\documentclass[prc,aps,a4paper,groupedaddress,superscriptaddress,nofootinbib,showpacs
,preprintnumbers,onecolumn]{revtex4}
\usepackage[english]{babel}
\usepackage{color}
\usepackage{graphicx}
\usepackage{amsfonts}
\usepackage{amssymb}
\usepackage{amsmath}
\usepackage{natbib}
\usepackage{dcolumn}
\usepackage{bm} 
\newcommand{\bwt}{\begin{widetext}}
\newcommand{\ewt}{\end{widetext}}
\newcommand{\beq}{\begin{equation}}
\newcommand{\eeq}{\end{equation}}
\newcommand{\bea}{\begin{eqnarray}}
\newcommand{\eea}{\end{eqnarray}}

\begin{document}
\def\e{{\rm e}}
\def\d{{\rm d}}
\def\vO{{\bf\Omega}}
\def\vM{{\bf M}}
\def\vQ{{\bf Q}}
\def\vP{{\bf P}}
\title{The 1D Heisenberg antiferromagnet model by the variation after projection method}
\author{Aziz Rabhi}
\email{rabhi@teor.fis.uc.pt} \affiliation{CFC, Departamento de F\'\i
sica, Universidade de Coimbra, 3004-516 Coimbra, Portugal}
\affiliation{LPMC-FST, Universit\'e de Tunis El-Manar, Campus
Universitaire, Le Belv\'ed\`ere-1060, Tunisia}
\author{Marta Brajczewska}\email{marta@teor.fis.uc.pt}
\affiliation{CFC, Departamento de F\'\i sica, Universidade de
Coimbra, 3004-516 Coimbra, Portugal}
\author{Peter Schuck}
\email{schuck@ipno.in2p3.fr} \affiliation{Institut de Physique
Nucl\'eaire, IN2P3-CNRS, Universit\'e Paris-Sud, F-91406 Orsay
Cedex, France} \affiliation{LPM2C, Maison des Magist\`eres-CNRS, 25,
av. des Martyrs, BP 166, 38042, Grenoble Cedex, France}
\author{Jo\~ao da Provid\^encia}
\email{providencia@teor.fis.uc.pt} \affiliation{CFC, Departamento de
F\'\i sica, Universidade de Coimbra, 3004-516 Coimbra, Portugal}
\author{Raouf Bennaceur}
\email{raouf.bennaceur@fst.rnu.tn} \affiliation{LPMC-FST,
Universit\'e de Tunis El-Manar, Campus Universitaire, Le
Belv\'ed\`ere-1060, Tunisia}
\date{\today}
\begin{abstract}

The 4 sites and 8 sites 1D anti-ferromagnetic Heisenberg chains in
the Jordan-Wigner representation are investigated within the
standard Hartree-Fock and RPA approaches, both in the symmetry
unbroken and in the symmetry broken phases. A translation invariant
groundstate, obtained by the projection method as a linear
combination of a symmetry-broken HF state and its image under
reflection, is also considered, for each chain type. It is found
that the projection method considerably improves the HF treatment
for instance as far as the groundstate energy is concerned, but also
with respect to the RPA energies. The results are furthermore
confronted with the ones obtained within so-called SCRPA scheme.


\end{abstract}
\maketitle

\section{Introduction}
The 1D Anti-Ferromagnetic Heisenberg Model (AFHM)~\cite{bet0} is a
prominent many body research field. It was the first many-body model
to be solved exactly by Bethe with his famous ansatz ~\cite{bet1}
and, as a solvable model, it has proved to be very useful for
testing and developing many body approaches ever since. In spite of
tremendous progress in the understanding of many aspects of the
model, it remains a very active field of interest and
research~\cite{mikes,boug}.

In \cite{rpp06} we have considered the anisotropic Heisenberg model
\begin{eqnarray}H=\sum_{n=1}^N\left(\displaystyle{1\over2}(S_{n+1}^+S_{n}^-+S_{n+1}^-S_{n}^+)
+gS_{n+1}^zS_{n}^z\right)\end{eqnarray} in the Jordan-Wigner
representation. We have found that for low values of the anisotropy
factor $g$, the Hartree-Fock (HF) groundstate is invariant under
translations. However, for large $g$ values a  (spherical) HF state
is no longer stable.  Instead one has to introduce a symmetry broken
(deformed) HF state which becomes degenerate. In \cite{rpp06}, a non
translational-invariant HF groundstate was introduced in order to
avoid unstable RPA modes, i. e., RPA modes with purely imaginary or
complex energies. However, it is well known that the exact
groundstate of the Heisenberg model is not degenerate (except for
$g=\infty$), so that the occurrence of a degenerate HF groundstate
is a spurious by product of the HF approximation. It is therefore
natural to look for an improved description by considering linear
combinations of deformed HF states. This is the main objective of
the present note. We are therefore lead to introduce a projection
method, implementing the Peierls-Yoccoz projection approach in the
version {\it variation after projection} \cite{RS}. As a result,
considerable improvement of  the HF treatment is achieved, not only
as far as the groundstate energy is concerned, but also with respect
to the RPA energies. We also give the results obtained with an
extension of RPA, the so-called SCRPA (self-consistent
RPA)\cite{rpp06}. It may also be observed that the projection method
results become exact for the 4-spin AFHM.

\section{Formalism}
\subsection{Variation after projection method}
We consider  the one dimensional Heisenberg antiferromagnet model~\cite{bet0} in
the Jordan-Wigner representation~\cite{JW}. In momentum space, the Hamiltonian
may be written~\cite{rpp06}
$$H={N\over4}+\sum_q\varepsilon^0_q\psi^{\dag}_q\psi_q+{g\over2}
\sum_{q_1q_2q_3q_4}
V_{q_1q_2q_3q_4}\psi_{q_1}^{\dag}\psi_{q_2}^{\dag}\psi_{q_4}\psi_{q_3}
\;
$$
where the $\psi^{\dag}_q, \psi_q$ are the usual fermion creation and
destruction operators with momentum $q$ and
$$\varepsilon^0_q=\cos q-1,\quad V_{q_1q_2q_3q_4}={1\over N}
(\cos(q_1-q_3)+\cos(q_2-q_4))\delta(q_1+q_2-q_3-q_4)
$$
with
$$q={j\pi\over N},\quad \left\{\begin{matrix}j=\pm1,\pm3,\cdots\pm(N-1),\hfill&\quad{\rm~if~} N/2 {\rm  ~is~even
}\\
j=0,\pm2,\pm4,\cdots \pm (N-2), N,\hfil&\quad{\rm ~if~} N/2 {\rm
~is~odd}\end{matrix}\right.
$$
We concentrate on chains with equal numbers of spins up and down, so
that $N/2$ is the number of fermions. The Hartree-Fock (HF) state is
obtained by filling up the lowest $N/2$ single particle levels. The
set of these levels may be written
$$D=\left\{q:\;{\pi\over2}< q\leq\pi\;\;{\rm or}\;\;-\pi\leq q<-{\pi\over2} 
\right\},
$$
so that the (translation invariant) HF state reads
$$| \Phi\rangle
=\prod_{h\in D}\psi_h^{\dag}|0\rangle.
$$
\begin{figure}[ht]
\vspace{1.5cm} \centering
\includegraphics[width=0.75\linewidth,angle=0]{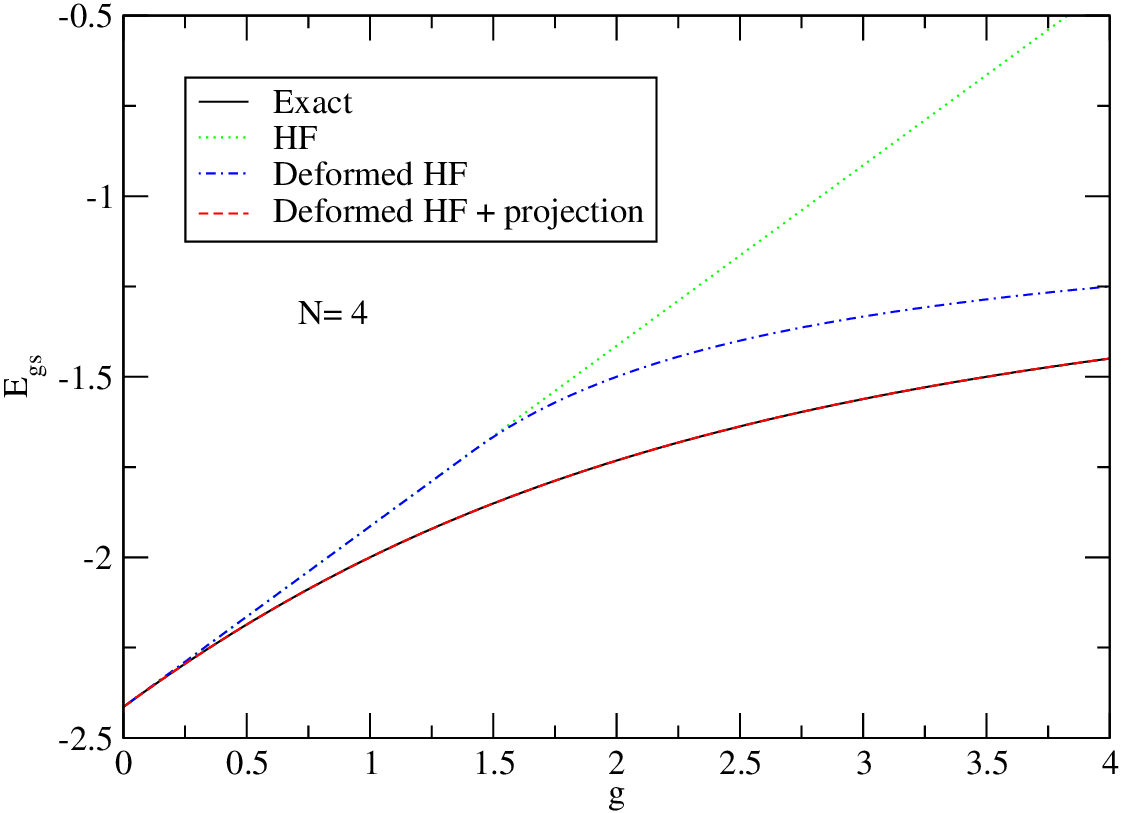}
\caption{(Color online) Four site system. Groundstate energy versus
$g$. The green dotted line shows the non-deformed HF result. The
blue dot-dashed line shows the deformed HF result. The red dashed
lines shows the projection method result (variation after
projection), superimposed on the exact result shown by the black
solid line.} \label{fig1}
\end{figure}
It has been shown in~\cite{rpp06} that  some frequencies of the RPA
modes associated with $|\Phi\rangle$, with momentum $q$ close to
$\pi$, (i.e., $q>q_c\approx 0.91\pi$), are complex, implying that
this state is unstable and other HF states exist which are
energetically preferred.  
We consider the canonical transformation which couples levels with
momenta $q$ and $q-\pi$,
$$
\zeta_{h}^{\dag}=\alpha_{h}
\psi_{h}^{\dag}+\beta_{h}\psi_{h-\pi}^{\dag}, \quad
\zeta_{h-\pi}^{\dag}=\alpha_{h}
\psi_{h-\pi}^{\dag}-\beta_{h}\psi_{h}^{\dag},\quad
h\in[{\pi\over2},\pi]\cup[-\pi,-{\pi\over2}]=D,
$$
where $\alpha_h,\beta_h,$ are real parameters such that
$\alpha_h^2+\beta_h^2=1$. This transformation breaks the
translational symmetry of the model. We say that the basis of the
states $\psi_q$ is spherical and the basis of the states $\zeta_q$
is deformed. It is well known that, very often, an energy gain may
be obtained in mean field approaches by breaking some symmetry of
the Hamiltonian. The HF state,
\begin{equation} |\widetilde \Phi\rangle =\prod_{h\in
D}\zeta_h^{\dag}|0\rangle,\label{tildePhi}
\end{equation} destroys translational invariance and leads to a lower energy
expectation value for some non-vanishing values of the parameters
$\alpha_k,\beta_k$, which should be determined variationally.
Indeed, let us consider the unitary translation operator
$$U_n=\e^{inP},\quad P=\sum_qq\psi_q^{\dag}\psi_q.
$$
We have
$$U_n\psi^{\dag}_qU_n^{\dag}=\e^{inq}\psi^{\dag}_q,\quad
U_n\zeta_{h}^{\dag}U_n^{\dag}=\e^{inh}(\alpha_{h}
\psi_{h}^{\dag}+\e^{-in\pi}\beta_{h}\psi_{h-\pi}^{\dag}).
$$
\begin{figure}[ht]
\centering
\includegraphics[width=0.75\linewidth,angle=0]{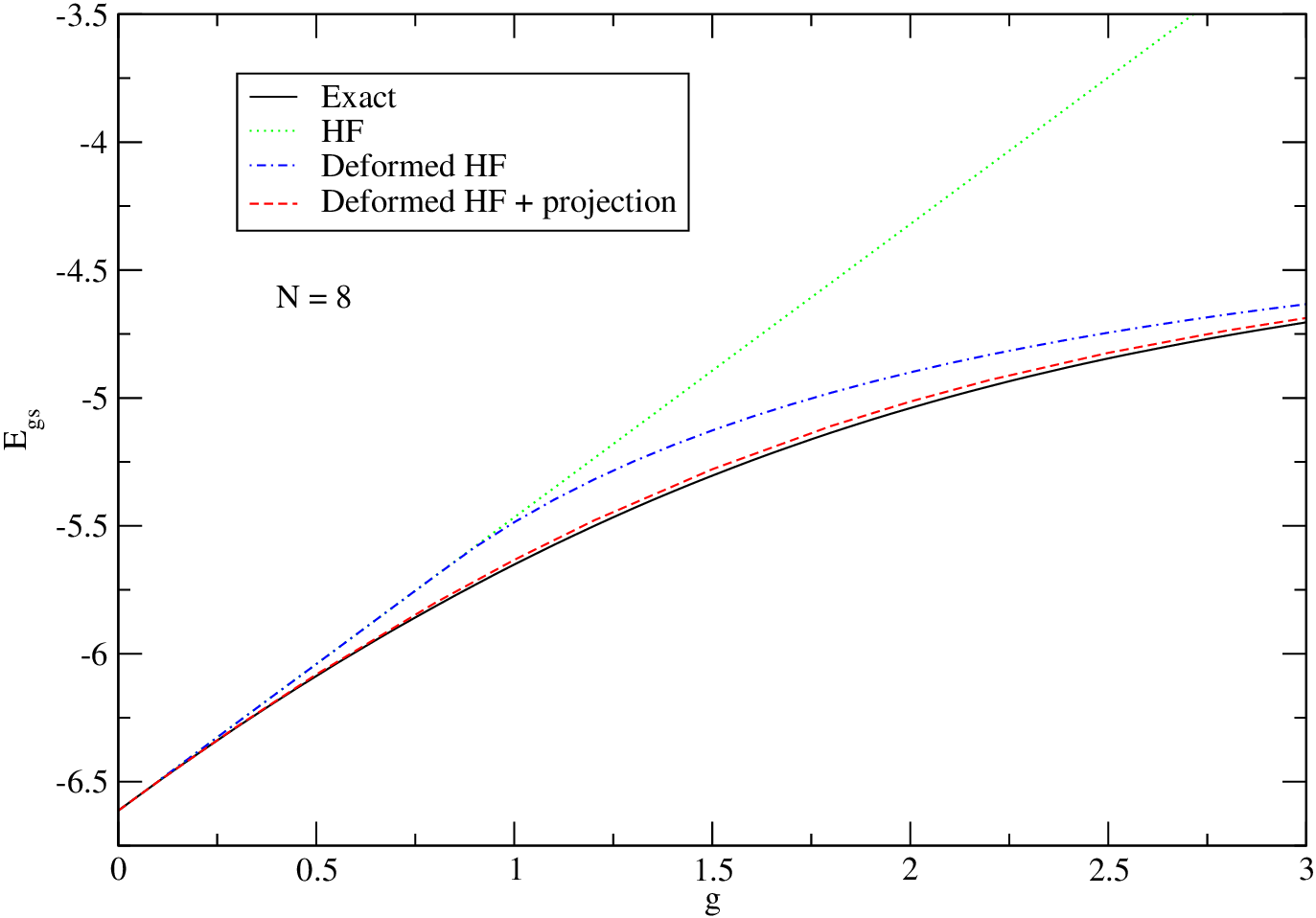}
\caption{(Color online) The same as Fig.~\ref{fig1}, but for the
eight site system (N=8).} \label{fig2}
\end{figure}
Thus, $| \Phi\rangle$ is an eigenstate of $U_n$, but $|\widetilde
\Phi\rangle$ is not. The deformed HF states $|\widetilde
\Phi_n\rangle=U_n|\widetilde \Phi\rangle,$ are not all  distinct.
There are only two distinct ones, for even and for odd $n$. Indeed,
$|\widetilde \Phi_0\rangle=|\widetilde \Phi_2\rangle=\cdots=
|\widetilde \Phi\rangle$ and $|\widetilde \Phi_1\rangle=|\widetilde
\Phi_3\rangle=\cdots=  U_1|\widetilde \Phi\rangle.$ As Figs.
\ref{fig1} and \ref{fig2} show, the groundstate HF energy, based on
the state $|\widetilde \Phi\rangle$, which breaks translational
symmetry, greatly improves the prediction of the symmetrical state
$|\Phi\rangle$. However, in the spirit of the Peierls-Yoccoz method
\cite{RS}, it is natural to look for a still improved description in
terms of a linear combination of the degenerate deformed states,
$|\widetilde \Phi\rangle$ and $|\widetilde \Phi_1\rangle$, which
amounts to considering the symmetrical projection of $|\widetilde
\Phi\rangle$. Since
$U_1(1+U_1)|\widetilde\Phi\rangle=(1+U_1)|\widetilde\Phi\rangle,$
the state $(1+U_1)|\widetilde\Phi\rangle$ is translation invariant.
It is an eigenstate of $P$ with zero eigenvalue.  We describe
briefly the computation of the expectation value of an arbitrary
many-body operator $X$ in the projected state,
$$
{\langle\widetilde\Phi|X(1+U_1)|\widetilde\Phi\rangle
\over\langle\widetilde\Phi|1+U_1|\widetilde\Phi\rangle}.$$ While
$|\widetilde\Phi\rangle$ is given by (\ref{tildePhi}), the reflected
state $U_1|\widetilde\Phi\rangle$ reads
 \begin{equation} U_1|\widetilde \Phi\rangle =\prod_{h\in
D}{\zeta'_h}^{\dag}|0\rangle=\prod_{h\in
D}U_1\zeta_h^{\dag}U_1^{\dag}|0\rangle.\label{UtildePhi}
\end{equation} We have
\begin{eqnarray*}&&U_1\zeta_h^{\dag}U_1^{\dag}={\zeta'_h}^\dag=\alpha_h\psi_h^\dag-\beta_h\psi_{h-\pi}^\dag
=(\alpha_h^2-\beta_h^2)\zeta_h^\dag-2\alpha_h\beta_h\zeta_{h-\pi}^\dag,\\
&&U_1\zeta_{h-\pi}^{\dag}U_1^{\dag}={{\zeta'}^\dag_{h-\pi}}=\alpha_{h}\psi_{h-\pi}^\dag+\beta_h\psi_{h}^\dag
=(\alpha_{h}^2-\beta_h^2)\zeta_{h-\pi}^\dag+2\alpha_h\beta_h\zeta_{h}^\dag.
\end{eqnarray*}
This implies that the particle-hole concept breaks down in
connection with the projected groundstate. The determination of
$\langle\widetilde\Phi|U_1|\widetilde\Phi\rangle$ and
$\langle\widetilde\Phi|XU_1|\widetilde\Phi\rangle$, where $X$ is
some operator, involves a large number of contractions with respect
to the vacuum $|0\rangle$. Obviously
$$\langle\widetilde\Phi|U_1|\widetilde\Phi\rangle=\prod_{h\in
D}\langle0|\zeta_h{\zeta'}^\dag_h|0\rangle=\prod_{h\in
D}(\alpha_h^2-\beta_h^2).$$ The computation of
$\langle\widetilde\Phi|XU_1|\widetilde\Phi\rangle$ is also
straightforward. We exemplify it for the case of a one-body
operator,
$X=\displaystyle\sum_{p_1,p_2}x_{p_1p_2}\zeta_{p_1}^\dag\zeta_{p_2},$
$$\langle\widetilde\Phi|XU_1|\widetilde\Phi\rangle=\sum_{h\in D}\left(\left(x_{hh}\alpha_h^2+(x_{h(h-\pi)}+x_{(h-\pi)h})\alpha_h\beta_h+
x_{(h-\pi)(h-\pi)}\beta_h^2\right)\prod_{h\neq k\in
D}(\alpha_k^2-\beta_k^2)\right).$$

The expectation value of the Hamiltonian in the projected state may
be written
\begin{equation}\label{calE0}{\cal E}_0(
\beta_{1},\cdots\beta_{{N/2}})={\langle\widetilde\Phi|H(1+U_1)|\widetilde\Phi\rangle
\over\langle\widetilde\Phi|1+U_1|\widetilde\Phi\rangle},\end{equation}
where we have explicitly indicated the dependence on the parameters
$\beta_{1},\cdots\beta_{{N/2}},$ with respect to which minimization
should be implemented in the end.

In Fig. \ref{fig1}, the performance of the three approaches, HF,
deformed HF and variation after projection are compared, for the 4
spin system, in connection with the estimation of the groundstate
energy. It is remarkable that the result obtained by the variation
after projection method, coincides with the exact result. In
Fig. \ref{fig2}, the corresponding results for the 8 spin system are
presented. Again, the performance of the projection state method is
remarkable.
\subsection{Random phase approximation method}
It is well known that the stability of the HF state $|\Phi\rangle$
may be investigated by studying the perturbed state
$|\Phi_S\rangle=\exp(iS)|\Phi\rangle$ where $S$ is an infinitesimal
one-body Hermitian operator. Expanding the expectation value of the
Hamiltonian, we have
$$\langle\Phi_S|H|\Phi_S\rangle=\langle\Phi|H|\Phi\rangle
+{1\over2}\langle\Phi|[S,[H,S]]|\Phi\rangle+\cdots,
$$
where, in the right hand side, a term
$-i\langle\Phi|[S,H]|\Phi\rangle$ has been omitted, since, by the HF
condition, it vanishes. Stability of $|\Phi\rangle$ means that
$\langle\Phi|[S,[H,S]]|\Phi\rangle$ is positive semi-definite.

The mean field dynamics, which is prescribed by the time evolution
of $|\Phi_S\rangle$, is determined by the quantal Lagrangian
$${\cal L}={i\over2}(\langle\Phi_S|\dot\Phi_S\rangle-\langle\dot\Phi_S
|\Phi_S\rangle)-\langle\Phi_S|H|\Phi_S\rangle,
$$
where the dot sign denotes time derivative, as usual. For small $S$,
we may replace ${\cal L}$ by its quadratic part,
$${\cal
L}^{(2)}={i\over2}\langle\Phi|[S,\dot
S]\Phi\rangle-{1\over2}\langle\Phi|[S,[H,S]]|\Phi\rangle,
$$
the time dependence being encapsulated in $S$. The time evolution of
$S$ is determined by the equation $${i}\langle\Phi|[\Xi,\dot
S]\Phi\rangle-\langle\Phi|[\Xi,[H,S]]|\Phi\rangle=0,
$$
where $\Xi$ denotes an arbitrary one-body operator (for instance,
$\Xi=\psi_p^\dagger\psi_q$). The ansatz $S=\exp(-i\omega t)\Theta$
leads to the equations of the so-called random phase approximation
(RPA),
$$\omega\langle\Phi|[\Xi,
\Theta]|\Phi\rangle-\langle\Phi|[\Xi,[H,\Theta]]|\Phi\rangle=0.
$$
If $\omega$ is not a real number, then 
$|\Phi\rangle$ is not stable and should be replaced by a stable
solution. This is what is done when the symmetrical state
$|\Phi\rangle$ is replaced by the deformed state
$|\widetilde\Phi\rangle$ for $g>g_c$, the critical strength for the
emergence of the instability. Although, in the conventional RPA,
$|\Phi\rangle$ is a Slater determinant, extended versions of the
RPA, for which $|\Phi\rangle$ is replaced by a state containing
correlations, have been proposed and  may be found in the literature
\cite{row,RS,scrp2}. In the case of the Heisenberg model in the
Jordan-Wigner representation, it is advantageous to replace
$|\Phi\rangle$ by the projected state
$(1+U_1)|\widetilde\Phi\rangle$, in the RPA equations. This is
permissible because, for $p\neq q$, we have
$$\langle\widetilde\Phi|(1+U_1)
[\psi_p^\dagger\psi_q,H](1+U_1)|\widetilde\Phi\rangle=0,$$ since the
projected state has vanishing momentum and the operator
$\psi_p^\dagger\psi_q$ carries a finite momentum $p-q$. The ket
$|\widetilde\Phi\rangle$ entering in the projected state should be
fixed by the variation after projection procedure \cite{RS}, that
is, the parameters $\beta_{1},\cdots,\beta_{{N/2}}$ should be
determined by minimizing ${\cal
E}_0(\beta_{1},\cdots,\beta_{{N/2}})$ given by (\ref{calE0}).

In order to determine the RPA frequencies, it is enough to consider
the ground state expectation values of the operators
\begin{align}{\cal
O}_N(q_2p_2;p_1q_1)&=[\psi_{p_2}^{\dag}\psi_{q_2},\psi_{p_1}^{\dag}\psi_{q_1}]=(\psi^{\dag}_{p_2}\psi_{q_1}\delta_{q_2p_1}-
\psi^{\dag}_{p_1}\psi_{q_2}\delta_{q_1p_2})\\ {\cal
O}_T(q_2p_2;p_1q_1)&=[\psi_{p_2}^{\dag}\psi_{q_2},[\sum_{a}\varepsilon_{a}\psi_{a}^{\dag}
\psi_{a},\psi_{p_1}^{\dag}\psi_{q_1}]]=(\varepsilon_{p_1}-\varepsilon_{q_1})(\psi^{\dag}_{p_2}\psi_{q_1}\delta_{q_2p_1}-
\psi^{\dag}_{p_1}\psi_{q_2}\delta_{q_1p_2})\end{align}
\begin{align}\label{OV}
{\cal O}_V(q_2p_2;p_1q_1)&=[\psi_{p_2}^{\dag}\psi_{q_2},[
{1\over2}\sum_{abcd}V_{abcd}\psi_{a}^{\dag}\psi_{b}^{\dag}
\psi_{d}\psi_{c},\psi_{p_1}^{\dag}\psi_{q_1}]]\nonumber\\ &
=2\sum_{bc}V_{q_2bcp_1}\psi^{\dag}_{p_2}\psi^{\dag}_b\psi_{q_1}\psi_c
+2\sum_{bc}V_{q_1bcp_2}\psi^{\dag}_{p_1}\psi^{\dag}_b\psi_{q_2}\psi_c
\nonumber\\&
-\sum_{ab}V_{abp_2p_1}\psi^{\dag}_{a}\psi^{\dag}_{b}\psi_{q_1}\psi_{q_2}
-\sum_{cd}V_{q_1q_2cd}\psi^{\dag}_{p_1}\psi^{\dag}_{p_2}\psi_{d}\psi_{c}
\nonumber\\&
-\sum_{bcd}V_{q_1bcd}\psi^{\dag}_{p_2}\psi^{\dag}_{b}\psi_{d}\psi_{c}\delta_{q_2p_1}
-\sum_{abd}V_{abp_1d}\psi^{\dag}_{a}\psi^{\dag}_{b}\psi_{d}\psi_{q_2}\delta_{q_1p_2}
. \end{align}
\begin{figure}[ht]
\vspace{1.5cm} \centering
\includegraphics[width=0.75\linewidth,angle=0]{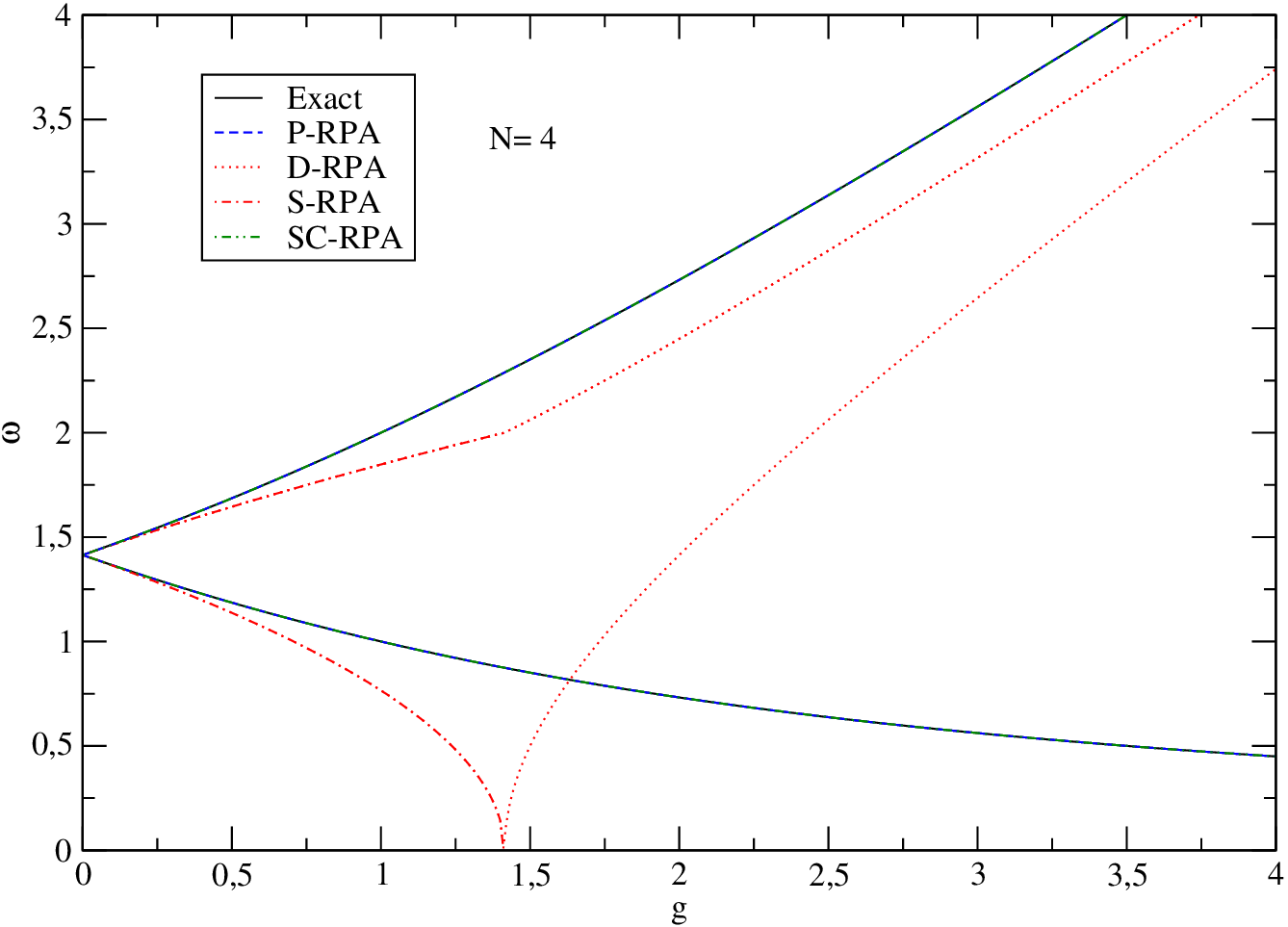}
\caption{(Color online) Four site system. RPA frequencies vs $g$.
The modes for momentum transfer $\displaystyle  q={\pi\over 2}$ are
degenerate with the upper mode for the momentum transfer
$\displaystyle q=\pi$. The RPA results are represented by the red
dot-dashed lines. The results for the projected state and SCRPA
shown by the blue dashed lines and the green dot-dashed lines,
respectively, coincide with the exact results which are shown by the
black full curves.} \label{fig3}
\end{figure}

It is convenient to consider in above operators $p_1-q_1=q_2-p_2$ what implies
momentum conservation. The
commutators and
double-commutators have been worked out in the symmetrical basis,
although the deformed basis is equally admissible for this purpose.
However, the expectation values defining the RPA matrices should be
taken in the appropriate state: the spherical HF state, the deformed
HF state or the projected state, sticking to the variation after
projection prescription.
The RPA frequencies are the eigenvalues $\omega$ of the eigenvalue
problem $$\omega{\cal N} \zeta=({\cal T}+g{\cal V})\zeta,$$ where
$\zeta$ denotes the eigenvector and ${\cal N}, {\cal T}, {\cal V}$
are appropriate matrices. Explicitly we have,
\begin{figure}[ht]
\vspace{1.5cm} \centering
\includegraphics[width=0.75\linewidth,angle=0]{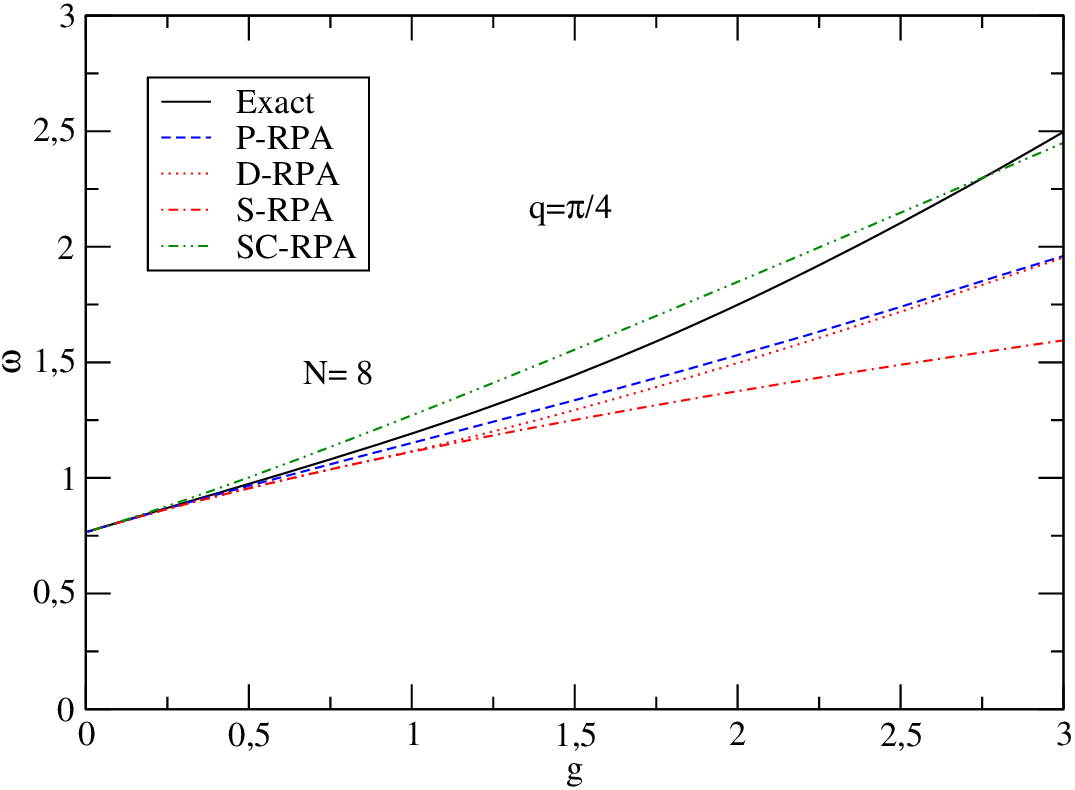}
\caption{(Color online) Eight site system. RPA frequencies vs
$g$
for the momentum transfer $\displaystyle q={\pi\over 4}$. The exact
results are shown by the black full curve.
The performances of S-RPA, D-RPA, P-RPA and SCRPA are compared.}
\label{exci4}
\end{figure}
$$\omega\sum_{p_1q_1}{\cal N}_{(q_2p_2),(p_1q_1)}
\zeta_{p_1q_1}=\sum_{p_1q_1}({\cal T}_{(q_2p_2),(p_1q_1)}+g{\cal
V}_{(q_2p_2),(p_1q_1)})\zeta_{p_1q_1},$$
where
\begin{align}{\cal N}_{(q_2p_2),(p_1q_1)}&=
\langle\widetilde\Phi|{\cal
O}_N(q_2p_2;p_1q_1)|\widetilde\Phi\rangle,\\ {\cal
T}_{(q_2p_2),(p_1q_1)}&=\langle\widetilde\Phi|{\cal
O}_T(q_2p_2;p_1q_1)|\widetilde\Phi\rangle,\\ {\cal
V}_{(q_2p_2),(p_1q_1)}&=\langle\widetilde\Phi|{\cal
O}_V(q_2p_2;p_1q_1)|\widetilde\Phi\rangle.
\end{align}
\begin{figure}[ht]
\centering
\includegraphics[width=0.75\linewidth,angle=0]{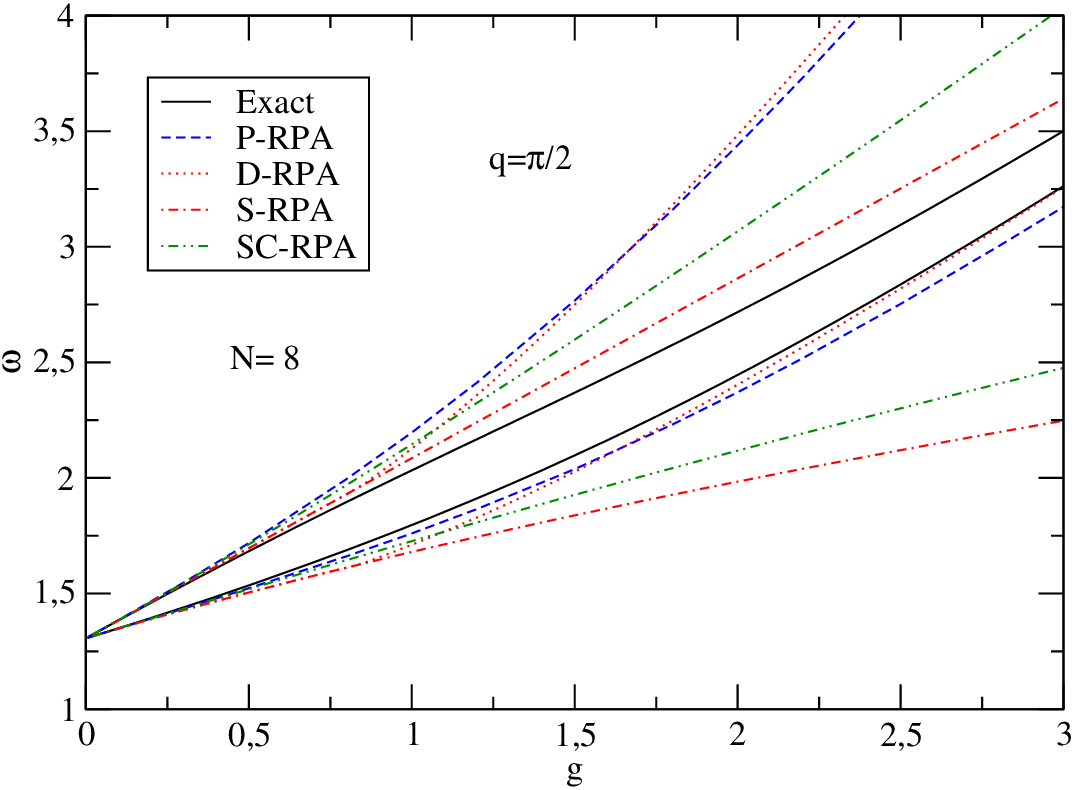}
 \caption{(Color online) Eight site system. RPA
frequencies vs
$g$
for the momentum transfer $\displaystyle q={\pi\over 2}$. The exact
results are shown by the  full curve in black color. The
performances of D-RPA, P-RPA and SC-RPA are compared. }
\label{exci2}
\end{figure}

The RPA frequencies based on the projected state are the eigenvalues
$\omega$ of the eigenvalue problem
$$\omega({\cal N}+{\cal N}') \zeta=(({\cal T}+{\cal T}')+g({\cal V}+{\cal V}'))\zeta,$$
or, explicitly,
$$\omega\sum_{p_1q_1}
({\cal N}+{\cal N}')_{(q_2p_2),(p_1q_1)}
\zeta_{p_1q_1}=\sum_{p_1q_1}(
({\cal T}+{\cal T}')_{(q_2p_2),(p_1q_1)}+g
({\cal V}+{\cal V}')_{(q_2p_2),(p_1q_1)})\zeta_{p_1q_1},$$ where the
matrices ${\cal N}', {\cal T}', {\cal V}'$ are given by
\begin{align}{\cal N}'_{(q_2p_2),(p_1q_1)}&= \langle\widetilde\Phi|{\cal
O}_N(q_2p_2;p_1q_1)U_1|\widetilde\Phi\rangle,\\ {\cal
T}'_{(q_2p_2),(p_1q_1)}&=\langle\widetilde\Phi|{\cal
O}_T(q_2p_2;p_1q_1)U_1|\widetilde\Phi\rangle,\\ {\cal
V}'_{(q_2p_2),(p_1q_1)}&=\langle\widetilde\Phi|{\cal
O}_V(q_2p_2;p_1q_1)U_1|\widetilde\Phi\rangle. \end{align}


The computation of the matrices ${\cal N}', {\cal T}', {\cal V}'$ is
straightforward, 
although somewhat more labor consuming than the computation of the
matrices ${\cal N}, {\cal T}, {\cal V}$. Some computational details
can be found in Appendix A.
\\

Besides the evaluation of RPA with the symmetry projected state, we
also show results corresponding to the so-called Self-Consistent RPA
(SCRPA) which was applied to the AFHM in our previous publication
\cite{rpp06}. The principle of this approach is explained in that
reference but for completeness, let us just outline the main aspects
of SCRPA. It is based on a particle-hole (ph) excitation operator of
the form

$$Q_{\mu}^{\dag} = \sum_{ph}[X^{\mu}_{ph}\psi^{\dag}_p\psi_h -
Y^{\mu}_{ph}\psi^{\dag}_h\psi_p]$$

\noindent where the excited state is given by $Q^{\dag}_{\mu}\vert 0
\rangle = \vert \mu \rangle$. The ground state should be the
vacuum to the corresponding destruction operator, that is
$Q_{\mu}\vert 0\rangle = 0$. Defining an average excitation energy
$\Omega_{\mu}$ via the energy weighted sum rule, i.e.

$$ \Omega_{\mu} = \frac{\langle 0\vert [Q_{\mu},[H,Q^{\dag}_{\mu}]]\vert 0\rangle}
{\langle 0\vert [Q_{\mu},Q^{\dag}_{\mu}]\vert 0\rangle}
$$ and minimizing with respect to the amplitudes $X, Y$, yields RPA
type of equations ${\langle 0\vert [\delta
Q_{\mu},[H,Q^{\dag}_{\mu}]]\vert 0\rangle} = \Omega_{\mu}{\langle
0\vert [\delta Q_{\mu},Q^{\dag}_{\mu}]\vert 0\rangle}$ with the
correlated ground state defined via the killing condition of above.
In the double commutator appear at most two body densities and also
one body densities which can be expressed as a functional of the $X,
Y$ amplitudes. Self consistent equations for the amplitudes $X, Y$
are then obtained which can be solved numerically by iteration.
More details can be found in ref \cite{rpp06}.

\section{Excitation energies: results and discussions}

\subsection{The four sites system}
Results for the four sites chain are summarized in Fig. \ref{fig3},
where we compare the excitation energies according to the spherical
RPA (S-RPA), the deformed RPA (D-RPA) and the RPA based on the
projected state (P-RPA)  with the exact excitation spectrum. For
$g<1.4$,  S-RPA shows some contact with the lower exact branch, but
this contact is completely lost by D-RPA, for $g>1.4$. However, the
excellent performance of P-RPA, which precisely reproduces the exact
spectrum, is remarkable.

\begin{figure}[ht]
\centering
\includegraphics[width=0.75\linewidth,angle=0]{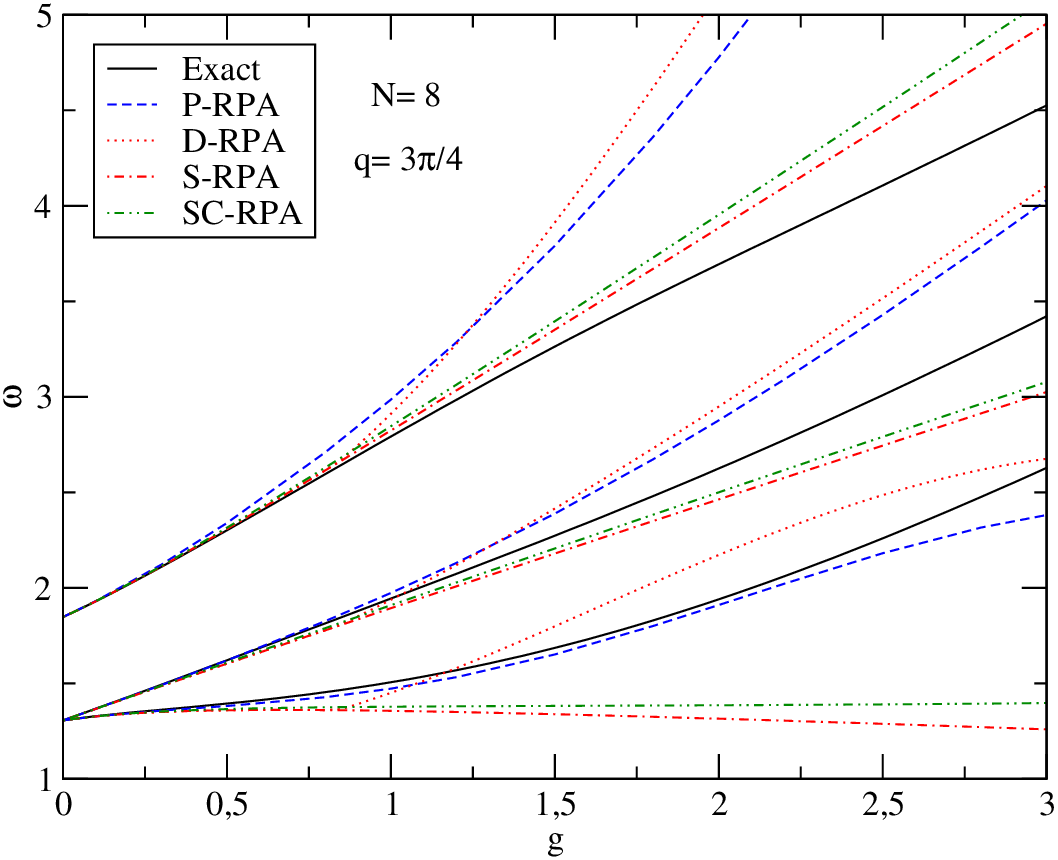}
\caption{(Color online) Eight site system. RPA frequencies vs $g$
for the momentum transfer $\displaystyle q={3\pi\over 4}$. The exact
results are shown by the black full curve, the D-RPA results by the
red dot-dashed lines, P-RPA by the blue dashed lines, and the SC-RPA
by the green dot-dashed lines.} \label{exci34}
\end{figure}

\begin{figure}[ht]
\vspace{1.5cm} \centering
\includegraphics[width=0.75\linewidth,angle=0]
{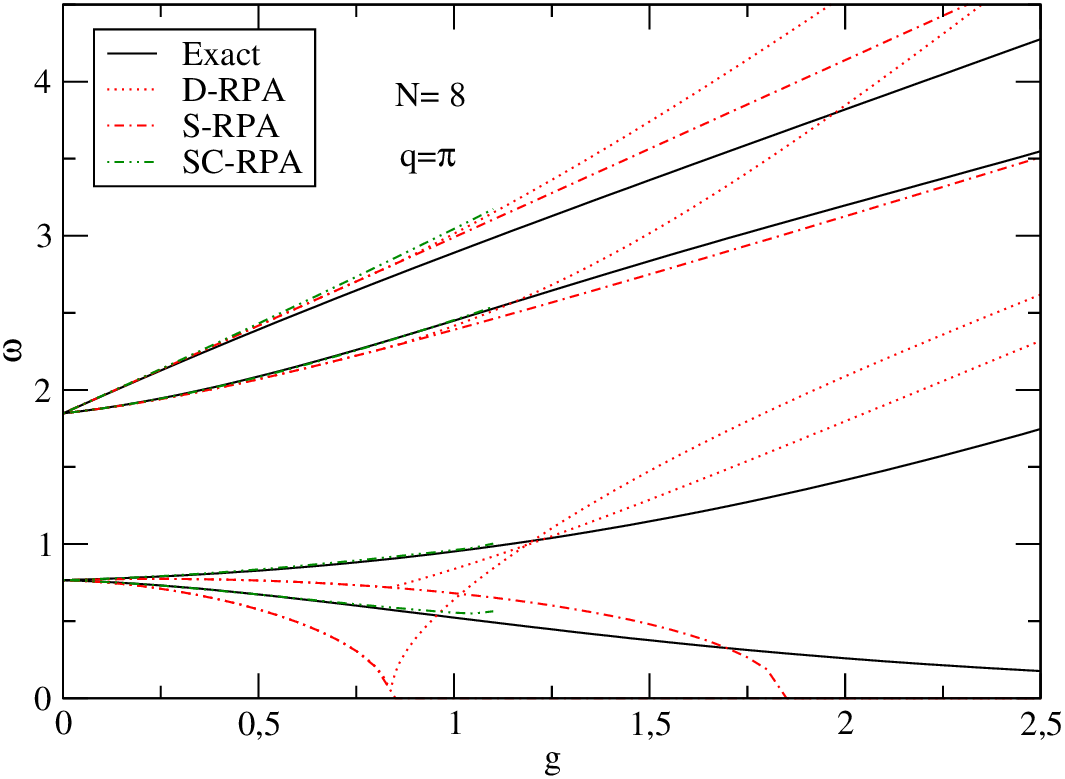} \caption{(Color online) Eight site system. RPA
frequencies vs $g$ for the momentum transfer $\displaystyle q=\pi$.
The exact results are shown by the black full curve, the S-RPA
results by the red dotted lines, the D-RPA results by the red
dot-dashed lines, the SCRPA results by the green dashed curves.}
\label{exci44}
\end{figure}
\begin{figure}[ht]
\vspace{1.5cm} \centering
\includegraphics[width=0.75\linewidth,angle=0]{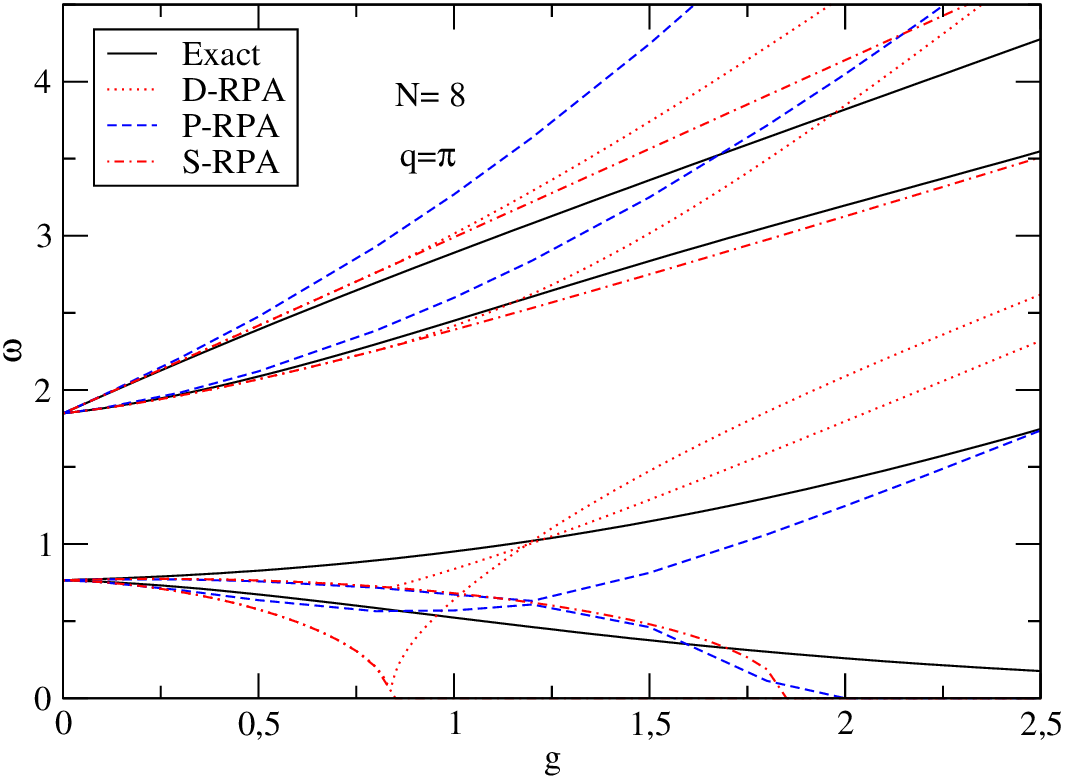}
\caption{(Color online) RPA frequencies vs $g$ for the momentum
transfer $\displaystyle q={\pi}$. The exact results are shown by the
full curve in black color. The performances of D-RPA, P-RPA and
SC-RPA are compared.
} \label{excipi}
\end{figure}
\subsection{The eight sites system}
We consider, now, the case of the eight sites chain. In figures
\ref{exci4}, \ref{exci2}, \ref{exci34}, \ref{exci44} and
\ref{excipi} we show the excitation spectrum as function of the
coupling constant $g$ (anisotropy factor), corresponding to the
momentum transfer $\displaystyle q=\frac{\pi}{4},
\frac{\pi}{2},\frac{3\pi}{4}$, and $\pi$, comparing the results of
S-RPA, D-RPA and P-RPA, with the exact ones. For $g \sim 0.7$, S-RPA
becomes unstable. Nevertheless, we represent it in Figs.
\ref{exci4}, \ref{exci2}, \ref{exci34}, \ref{exci44} for comparison
and because in some cases it appears, unexpectedly, to describe
better the physical situation than D-RPA. \color{black} In Fig.
\ref{exci4}, we notice that the performances of SCRPA, P-RPA and
D-RPA, which for $g<0.7$ goes over to the S-RPA, are of similar
quality, although P-RPA slightly improves over D-RPA. However,
beyond $g \sim 1$, D-RPA and also P-RPA depart significantly from
the exact solution. Only SCRPA stays reasonably close for large
values of $g$. The latter is somewhat surprising, since SCRPA is
here evaluated only in the `spherical' single particle basis which,
in the HF approximation, stops to be valid beyond $g \sim 0.7$ (as
will be seen later in Figs. \ref{exci44} and \ref{excipi}).

One may conclude that in what concerns the momentum transfer
$q=\pi/4$, all approaches, S-RPA, D-RPA, P-RPA perform reasonably
well below $g < 1$. The more reasonable agreement of SCRPA with the
exact values for $g > 1$ may be accidental. Surprisingly P-RPA does
not improve much on D-RPA. We will see that this is not always the
case for the other transfers and that quantum fluctuations
apparently are of different importance in the different channels. In
Fig. \ref{exci2}, for $q=\pi/2$, we see that the lowest state is
quite well described by D-RPA as well as by P-RPA indicating again
that quantum fluctuations are not very pertinent. On the other hand,
the second excited state is not well reproduced by any one of the
four approximations. Presumably this state has strong admixures of
the multi-particle multi-hole type. In Fig. \ref{exci34}, for
$q=3\pi/4$, we see for the first excited state a clear improvement
of P-RPA over D-RPA. Again the other two states are not very well
reproduced by any of the four approaches.

Because of its particular structure, we split the $q=\pi$ excited
state case into two figures. We first compare, in Fig. \ref{exci44},
the S-RPA and the D-RPA results with the exact one. It is clear that
there is no tendency for the D-RPA to reproduce the lowest exact
level, which means that this level is not of the particle-hole type
but of a different nature. Indeed, it is natural to assume that the
existence of two HF states is responsable for a tunneling mechanism
which splits the groundstate into a nearly degenerate doublet. In
principle, above $g_c\sim0.7$, S-RPA looses its meaning and one
might tend to ignore it. However, interestingly enough, the lowest
real S-RPA level, which becomes complex for $g>1.8$, reappears under
the guise of the lowest level of the P-RPA, as Fig. \ref{excipi}
shows. We also show in Fig. \ref{exci44} the results of SCRPA which
performs remarkably well in the spherical region (and beyond) for
all four states.

In Fig.~\ref{excipi}, which also refers to $q=\pi$, we observe that
the results of P-RPA improve the corresponding ones for the D-RPA
for the two lowest states. However, for the two upper states, the
situation is just the inverse: the P-RPA results are worse than the
D-RPA ones. In this respect we should be aware of the well known
fact that RPA overestimates the correlations and, therefore with
attractive interaction the states come at lower energies than with
an improved RPA. One sees this clearly for the two lowest states in
Fig. 8 where the additional repulsion coming from the projection
pushes the states upwards. The same upward push happens for the two
upper states and the reasonable agreement of the D-RPA results with
the exact solution must be interpreted as just an accident due to
too much correlations in D-RPA. In the deformed region both D-RPA
and P-RPA strongly deviate from the exact solution indicating that
higher than ph-correlations are contained in those states in the
deformed region. For $g>1.7$, the lowest exact state is clearly not
of the ph type.

It is remarkable that the P-RPA leads to a
level at almost zero excitation, for a surprisingly extended range
of $g$ values, mimicking a low lying state with odd parity which
quickly becomes nearly degenerate with the groundstate. In the next
subsection it is shown that this level is very well reproduced by
the  Peierls-Yoccoz projection method in the version variation after
projection \cite{RS}.

Let us finally mention that the states
$\psi^\dagger_{p_i}\psi_{q_i}~i=1,2,$ used for constructing the RPA
matrices $\cal N$, $\cal T$, $\cal V$, $\cal N'$, $\cal T'$, $\cal
V'$, were of the particle-hole type, that is, either $p_i\in
D,~q_j\not\in D$ or $p_i\not\in D,~q_j\in D$.

\subsection{The lowest excited state and the Peierls-Yoccoz method}

The  Peierls-Yoccoz projection method has been devised to describe
the rotational spectrum of a deformed nucleus. Let us assume that a
prolate ellipsoidal nucleus rotates around an axis perpendicular to
the symmetry axis. Let $|\Phi(\Omega)\rangle$ denote the wave
function of the rotated nucleus by an angle $\Omega$. According to
the Peierls-Yoccoz method, the wave function of the nucleus is a
superposition of the wave functions $|\Phi(\Omega)\rangle$,
$$|\Psi\rangle=\int \d\Omega f(\Omega)|\Phi(\Omega)\rangle,$$(see
Ref. \cite{RS}). From the symmetry of the nuclear Hamiltonian, it
follows that $f(\Omega)={\rm exp}(im\Omega)$. For $m=0$, a
description of the groundstate is obtained. For $ m=1,2,\ldots$ a
rotational band arises.  In the Heisenberg model, the analogous
symmetry to rotational symmetry is reflection symmetry, which is
broken by the deformed HF state, so that we are dealing with a
discrete group and the integral is replaced by a sum
$|\Psi\rangle=(f_0+f_1 U_1)|\tilde\Phi\rangle$. From the symmetry
properties of the Hamiltonian it follows that either $f_0=f_1$
(groundstate, with even parity) or $f_0=-f_1$ (excited state with
odd parity). In this model, instead of a rotational spectrum, we
find a doublet of states. We consider the variation after projection
version of the method \cite{RS}.

We have already discussed the ratio
$${\langle\tilde\Phi|H(1+U_1)|\tilde\Phi\rangle\over\langle\tilde\Phi|1+U_1|\tilde\Phi\rangle }.$$
It is, therefore, suggestive that we also consider the ratio
$${\langle\tilde\Phi|H(1-U_1)|\tilde\Phi\rangle\over\langle\tilde\Phi|1-U_1|\tilde\Phi\rangle }$$ which may describe the lowest state of odd parity.
If we plot the difference
$$\delta={\langle\tilde\Phi|H(1-U_1)|\tilde\Phi\rangle\over\langle\tilde\Phi|1-U_1|\tilde\Phi\rangle }-
{\langle\tilde\Phi|H(1+U_1)|\tilde\Phi\rangle\over\langle\tilde\Phi|1+U_1|\tilde\Phi\rangle
}$$versus the coupling constant $g$, the curve which is obtained
falls almost on top of the lowest exact excited state in fig. 6. In
table I, we give $\delta$ as a function of the coupling strength
$g$, for the eight sites system.

The feature that both states get degenerate in the infinite $g$
limit is reproduced by the deformed HF state. Indeed, if we denote
by $\phi_k^\dagger$ the creation operator of a fermion in site $k$
($k\in\{1,2,3,4,5,6,7,8\}$), then the groundstate wave function is
two-fold degenerate and reads
$\phi_1^\dagger\phi_3^\dagger\phi_5^\dagger\phi_7^\dagger|0\rangle$
or
$\phi_2^\dagger\phi_4^\dagger\phi_6^\dagger\phi_8^\dagger|0\rangle$.
Now, in momentum representation we have
\begin{eqnarray*}&&\phi_k^\dagger={1\over\sqrt{8}}\left(\psi_{1}\e^{i{k\pi\over8}}+\psi_{-1}\e^{-i{k\pi\over8}}
+\psi_{3}\e^{i{3k\pi\over8}}+\psi_{-3}\e^{-i{3k\pi\over8}}\right.\\&&+\left.
\psi_{5}\e^{i{5k\pi\over8}}+\psi_{-5}\e^{-i{5k\pi\over8}}
+\psi_{7}\e^{i{7k\pi\over8}}+\psi_{-7}\e^{-i{7k\pi\over8}}\right)\\&&
={1\over\sqrt{8}}\left(\e^{i{k\pi\over8}}(\psi_{1}+\psi_{-7}\e^{-i{k\pi}})
+\e^{-i{k\pi\over8}}(\psi_{-1}+\psi_{7}\e^{i{k\pi}})\right.\\&&+\left.
\e^{i{3k\pi\over8}}(\psi_{3}+\psi_{-5}\e^{-i{k\pi}})
+\e^{-i{3k\pi\over8}}(\psi_{-3}+\psi_{5}\e^{i{k\pi}})\right),\end{eqnarray*}
implying that the deformed HF becomes exact in the limit
$g\rightarrow\infty$. In this limit, one can neglect the kinetic
energy in (1) and one may check that the above state is eigenstate
to the interaction part of (1) alone. 
\begin{table}
\caption{\small{Description of the lowest excited state by the P-RPA
and by the Peierls-Yoccoz projection}}
\begin{ruledtabular}
\begin{tabular}{ c c c c}
 g        & $\omega_1$(Exact)  & $\omega_1$ (P-RPA) & $\delta$ \\
\hline
0.0 & 0.765367 & 0.765367 & 0.765367 \\
0.5 & 0.672780 & 0.636131 & 0.666271 \\
1.0 & 0.522674 & 0.568805 & 0.504778 \\
1.5 & 0.375722 & 0.460095 & 0.350829 \\
2.0 & 0.258764 & 0.000000 & 0.233469 \\
2.5 & 0.176111 & 0.000000 & 0.154336 \\
3.0 & 0.121093 & 0.000000 & 0.103791 \\
\end{tabular}
\end{ruledtabular}
\label{tab}
\end{table}

\color{black}
\section{Conclusions, discussion, and outlook}
 We have investigated the 8 sites and the 4 sites 1D AFH model with HF and
RPA type of theories. One may wonder why mean field type of theories
should be applied to 1D systems. However, the 1D character only
becomes specific getting close to the thermodynamic limit. As long
as the dimensions of the discrete RPA problem stay rather low,i.e.,
the number of sites relatively small, whether one works in 3D or in
1D has no real significance. This can, e.g., be seen in the case of
other 1D lattice models of the Hubbard type where RPA kind of
approaches give very encouraging results, see for example
\cite{jem,baril}. As is well known RPA may signal instabilities of
the system. In order to avoid unstable RPA modes for large values of
the anisotropy factor $g$, the symmetry unbroken HF state, i.e. the
translational invariant HF state, defined in terms of a symmetrical
single particle basis, should and has been replaced by a
symmetry-broken HF state, defined in terms of a deformed single
particle basis. Since the deformed HF state is degenerate, it is
natural to apply the Peierls-Yoccoz projection method, in the
version variation after projection \cite{RS}. \color{black} When the
projection method is used, the groundstate is described as a
superposition of two linearly independent deformed HF states and the
deformed HF basis is shown to play a role even for small $g$ values,
for which the symmetrical HF state is stable. A formalism of the RPA
type based on the projected groundstate, was outlined. The projected
RPA has in common with the self-consistent RPA (SCRPA) \cite{scrp2}
the fact that it is not related to a specific HF state, but rather
to a correlated groundstate. However, the similarity between that
theory and the present approach may stop here. We have found that
the projection method considerably improves the HF treatment mostly
as far as the groundstate energy is concerned, but also with respect
to the lower RPA energies. While in D-RPA low lying levels are
absent for $g>0.8$, in P-RPA a low lying level which quickly
approaches 0 when $g\rightarrow\infty$, is found. The existence of
this level is interesting, even though  it deviates from the actual
first excited state, which becomes degenerate with the ground state
in $g\rightarrow\infty$ limit. Tunneling processes, which become
inhibited in this limit, are poorly described by P-RPA and are
ignored by D-RPA. We also observe that the projection method results
become exact for the 4-site AFHM. In the 8-site problem, we applied
the projection technique for both ground states, i.e. for positive
parity and negative parity obtaining thus the first excited state
with high accuracy.

Concerning future developments, it might be interesting to formulate
SCRPA also with a deformed single particle basis and apply the
projection technique on top of it. Whether such an extension of the
present approach will be
feasible remains to be seen.

\section*{Acknowledgements}
A.R. acknowledges kind hospitality of University of Coimbra, where
most of this work was done.

\section*{Appendix A 
}

As an example, we present some computational details for the 4-site
system.

The HF state reads
$|\Phi\rangle=\psi_{h_1}^{\dag}\psi_{h_2}^{\dag}|0\rangle,\;h_1={3\over4}\pi,\;h_2=-{3\over4}\pi$.
The deformed HF state
$$|\tilde\Phi\rangle=\zeta_{h_1}^{\dag}\zeta_{h_2}^{\dag}|0\rangle,
\quad\zeta_{h_i}^{\dag}=\alpha\psi_{h_i}^{\dag}-\beta\psi_{h_i+\pi}^{\dag},\quad
i=1,2,\quad\alpha,\beta\in {\bf R},\quad\alpha^2+\beta^2=1,
$$ is determined by minimizing, with respect to $\beta$, the
expectation value $$\langle\tilde\Phi|H|\tilde\Phi\rangle
=-1-(\alpha^2-\beta^2)\sqrt{2}+{g\over2}(\alpha^2-\beta^2)^2.
$$
 The variation after projection procedure is implemented  when we
minimize, with respect to $\beta$, the expression
\begin{equation}{\cal
E}_{Proj}={\langle\tilde\Phi|H(1+U_1)|\tilde\Phi\rangle
\over\langle\tilde\Phi|1+U_1|\tilde\Phi\rangle}
=-1+{-2(\alpha^2-\beta^2)\sqrt{2}+g(\alpha^2-\beta^2)^2\over1
+(\alpha^2-\beta^2)^2},\quad U_1=\e^{iP}.\label{EProj}\end{equation}
The minimum of ${\cal E}_{Proj}$ occurs for
$$\alpha^2-\beta^2={-g+\sqrt{8+g^2}\over\sqrt{8}}.
$$
For momentum transfer $\pm \pi/2$, we find  \begin{align}{\cal
N}&=(\alpha^2-\beta^2)\begin{pmatrix}1&0\\0&-1\end{pmatrix}\\
{\cal
T}&=-\sqrt{2}(\alpha^2-\beta^2)\begin{pmatrix}1&0\cr0&1\end{pmatrix}\\
{\cal V} &=-{g\over2}\begin{pmatrix}1&-1\\-1&1 \end{pmatrix}
\end{align}
 For the projected state we still need the matrices ${\cal
N}',{\cal T}',{\cal V}'$. We find ${\cal N}'={\cal N},{\cal
T}'={\cal T},{\cal V}'={\cal V}$.

For momentum transfer $\pi$, we find  \begin{align}{\cal
N}&=(\alpha^2-\beta^2){\rm diag}(-1,-1,1,1)
\\
{\cal T}&=\sqrt{2}(\alpha^2-\beta^2) {\rm diag}(1,1,1,1) \end{align}

\begin{equation}{\cal
V}=\begin{pmatrix}2\alpha^2\beta^2&-{1\over2}(\alpha^2-\beta^2)^2&-2{\alpha^2\beta^2}&{1\over2}(\alpha^2-\beta^2)^2
\\-{1\over2}(\alpha^2-\beta^2)^2&2\alpha^2\beta^2&{1\over2}(\alpha^2-\beta^2)^2&-2{\alpha^2\beta^2}\\
-2\alpha^2\beta^2&{1\over2}(\alpha^2-\beta^2)^2&2\alpha^2\beta^2&-{1\over2}(\alpha^2-\beta^2)^2\\
{1\over2}(\alpha^2-\beta^2)^2&-2\alpha^2\beta^2&-{1\over2}(\alpha^2-\beta^2)^2&2\alpha^2\beta^2\end{pmatrix}
\end{equation} For the projected state RPA we still need the matrices ${\cal
N}',~{\cal T}',~{\cal V}'$. We find ${\cal N}={\cal N}',~{\cal
T}={\cal T}'$ and
\begin{align}{\cal V}'&=\begin{pmatrix}0&-{1\over2}&0&{1\over2}
\\-{1\over2}&0&{1\over2}&0\cr 0&{1\over2}&0&-{1\over2}\\
{1\over2}&0&-{1\over2}&0\end{pmatrix}
\end{align}
The RPA frequencies based on the projected state are the roots of
$${\rm det}[({\cal N}+{\cal N}')\omega -({\cal T}+{\cal T}')-g({\cal V}+{\cal
V}')]=0.
$$
The parameter $\beta$ should be such that it minimzes ${\cal E}_0$
given by (\ref{EProj}).
\section*{Appendix-B}
We illustrate the determination of the matrices ${\cal N},{\cal
T},{\cal V},{\cal N}',{\cal T}',{\cal V}'$ for the 4 sites case. The
fermion operators read
$\psi_{3\pi\over4},\psi_{-{3\pi\over4}},\psi_{\pi\over4},\psi_{-{\pi\over4}},$
but, for simplicity of notation, we write
$\psi_{3},\psi_{-{3}},\psi_{1},\psi_{-{1}}$. The deformed fermion
operators read $\zeta_3=\alpha\psi_3+\beta\psi_{-1}$,
$\zeta_{-3}=\alpha\psi_{-3}+\beta\psi_{1}$,
$\zeta_1=\alpha\psi_1-\beta\psi_{-3}$,
$\zeta_{-1}=\alpha\psi_{-1}-\beta\psi_{3}$.  For momentum transfer
$q=\pi/2$, the matrix ${\cal N}$ reads
$${\cal N}=\begin{pmatrix}{\cal N}_{1,3;1,3}&{\cal N}_{1,3;3,1}\\{\cal N}_{3,1;1,3}&{\cal N}_{3,1;3,1}\end{pmatrix},$$
 where ${\cal
N}_{q_2p_2;p_1q_1}=\langle\widetilde\Phi|{\cal
O}_N(q_2p_2;p_1q_1)|\widetilde\Phi\rangle$, with
$(q_2p_2),(p_1q_1)\in\{(1,3),(3,1)\}$, being
$|\widetilde\Phi\rangle=\zeta_3^\dagger\zeta_{-3}^\dagger|0\rangle.$
The matrix ${\cal N}'$ reads
$${\cal N}'=\begin{pmatrix}{\cal N}'_{1,3;1,3}&{\cal N}'_{1,3;3,1}\\{\cal N}'_{3,1;1,3}&{\cal N}'_{3,1;3,1}\end{pmatrix},
$$
 where ${\cal
N}'_{q_2p_2;p_1q_1}=\langle\widetilde\Phi|{\cal
O}_N(q_2p_2;p_1q_1)U_1|\widetilde\Phi\rangle$. The corresponding
matrices ${\cal T},~{\cal V}$, ${\cal T}',~{\cal V}'$ are similarly
constructed.

For momentum transfer $q=\pi$, the matrix ${\cal N}$ reads
$${\cal N}=\begin{pmatrix}{\cal N}_{1,-3;1,-3}&{\cal N}_{1,-3;-1,3}&{\cal N}_{1,-3;-3,1}&{\cal N}_{1,-3;3,-1}\\
{\cal N}_{-1,3;1,-3}&{\cal N}_{-1,3;-1,3}&{\cal
N}_{-1,3;-3,1}&{\cal N}_{-1,3;3,-1}\\
{\cal N}_{-3,1;1,-3}&{\cal N}_{-3,1;-1,3}&{\cal N}_{-3,1;-3,1}&{\cal N}_{-3,1;3,-1}\\
{\cal N}_{3,-1;1,-3}&{\cal N}_{3,-1;-1,3}&{\cal N}_{3,-1;-3,1}&{\cal
N}_{3,-1;3,-1}\end{pmatrix},$$ where ${\cal
N}_{q_2p_2;p_1q_1}=\langle\widetilde\Phi|{\cal
O}_N(q_2p_2;p_1q_1)|\widetilde\Phi\rangle$,  with
$(q_2p_2),(p_1q_1)$ $\in\{(1,-3),(-1,3),(-3,1),(3,-1)\}$. 
The matrix ${\cal N}'$ reads
$${\cal N}'=\begin{pmatrix}{\cal N}'_{1,-3;1,-3}&{\cal N}'_{1,-3;-1,3}&{\cal N}'_{1,-3;-3,1}&{\cal N}'_{1,-3;3,-1}\\
{\cal N}'_{-1,3;1,-3}&{\cal N}'_{-1,3;-1,3}&{\cal
N}'_{-1,3;-3,1}&{\cal N}'_{-1,3;3,-1}\\
{\cal N}'_{-3,1;1,-3}&{\cal N}'_{-3,1;-1,3}&{\cal N}'_{-3,1;-3,1}&{\cal N}'_{-3,1;3,-1}\\
{\cal N}'_{3,-1;1,-3}&{\cal N}'_{3,-1;-1,3}&{\cal
N}'_{3,-1;-3,1}&{\cal N}'_{3,-1;3,-1}\end{pmatrix},$$ where ${\cal
N}'_{q_2p_2;p_1q_1}=\langle\widetilde\Phi|{\cal
O}_N(q_2p_2;p_1q_1)U_1|\widetilde\Phi\rangle$. The corresponding
matrices ${\cal T},~{\cal V}$, ${\cal T}',~{\cal V}'$ are similarly
constructed.

Next, we exemplify the computation of the entries ${\cal
V}_{1,-3;1,-3}$ and  ${\cal V}'_{1,-3;1,-3}$. From (\ref{OV}) we
find
\begin{eqnarray*}{\cal
O}_V(1,-3;1,-3)&=&4~v_{-3,1,-3,1}~\psi_{-3}^\dagger\psi_{1}^\dagger\psi_{1}\psi_{-3}
\\&-&4~v_{3,-3,1,-1}~\psi_{3}^\dagger\psi_{-3}^\dagger\psi_{-1}\psi_{1}~-4~v_{-3,-1,3,1}\psi^\dagger_{-3}\psi^\dagger_{-1}\psi_1\psi_3
\\&+&(2~v_{-1,1,-1,1}-2~v_{-3,-1,-3,-1})~(\psi_{-3}^\dagger\psi_{-1}^\dagger\psi_{-1}\psi_{-3}-\psi_{1}^\dagger\psi_{-1}^\dagger\psi_{-1}\psi_{1})\\
&+&(2~v_{-3,3,-3,3}-2~v_{1,3,1,3})~(\psi_{1}^\dagger\psi_{3}^\dagger\psi_{3}\psi_{1}-\psi_{-3}^\dagger\psi_{3}^\dagger\psi_{3}\psi_{-3})
.\end{eqnarray*} For instance, for the expectation value
$\langle\widetilde\Phi|\psi_{-1}^\dagger\psi_{-3}^\dagger\psi_{3}\psi_{1}|\widetilde\Phi\rangle
=\langle0|\zeta_3\zeta_{-3}\psi_{-1}^\dagger\psi_{-3}^\dagger\psi_{3}\psi_{1}\zeta^\dagger_{-3}\zeta_3^\dagger|0\rangle$
we obtain $$\langle0|\zeta_{-3}\psi_{-3}^\dagger|0\rangle
\langle0|\zeta_3\psi_{-1}^\dagger|0\rangle\langle0|\psi_{1}\zeta_{-3}^\dagger|0\rangle
\langle0|\psi_{3}\zeta^\dagger_{3}|0\rangle=-\alpha^2\beta^2.$$ On
the other hand, the expectation value
$\langle\widetilde\Phi|\psi_{-1}^\dagger\psi_{-3}^\dagger\psi_{3}\psi_{1}U_1|\widetilde\Phi\rangle$
$
=\langle0|\zeta_3\zeta_{-3}\psi_{-1}^\dagger\psi_{-3}^\dagger\psi_{3}\psi_{1}{\zeta'}^\dagger_{-3}{\zeta_3'}^\dagger|0\rangle$
becomes $$\langle0|\zeta_{3}\psi_{-1}^\dagger|0\rangle
\langle0|\zeta_{-3}\psi_{-3}^\dagger|0\rangle\langle0|\psi_{1}{\zeta'}_{-3}^\dagger|0\rangle
\langle0|\psi_{3}{\zeta'}^\dagger_{3}|0\rangle=\alpha^2\beta^2.$$
Finally, $\langle\widetilde\Phi|{\cal
O}_V({1,3;1,3})|\widetilde\Phi\rangle=2\alpha^2\beta^2$, while
$\langle\widetilde\Phi|{\cal
O}_V({1,3;1,3})U_1|\widetilde\Phi\rangle=0$\\\\
\end{document}